# Transformers Applied to Short-term Solar PV Power Output Forecasting

Andea Scott
Stanford Univeristy
andea98@stanford.edu

Folasade Ayoola
Stanford Univeristy
fayoola@stanford.edu

Sindhu Sreedhara
Stanford Univeristy
sindhus@stanford.edu

## Abstract

*Reliable forecasts of the power output from variable renewable energy generators like solar photovoltaic systems are important to balancing load on real-time electricity markets and ensuring electricity supply reliability. However, solar PV power output is highly uncertain, with significant variations occurring over both longer (daily or seasonally) and shorter (within minutes) timescales due to weather conditions, especially cloud cover.*

*This paper builds on existing work that uses convolutional neural networks in the computer vision task of predicting (in a Nowcast model) and forecasting (in a Forecast model) solar PV power output (Stanford EAO SUNSET Model). A pure transformer architecture followed by a fully-connected layer is applied to one year of image data with experiments run on various combinations of learning rate and batch size.*

*We find that the transformer architecture performs almost as well as the baseline model in the PV output prediction task. However, it performs worse on sunny days.*

## 1. Introduction

Electricity generation currently accounts for about a quarter of all greenhouse gas (GHG) emissions both in the US and globally [1][2]. Renewable electricity from solar photovoltaic (PV) cells can facilitate the decarbonization of this sector. However, the solar resource is intermittent in supply, since its output depends on the time of day and local weather conditions. The power output from a PV cell can drop by up to 80% in the span of a few minutes if clouds block solar radiation from reaching the PV array [3]. This phenomenon is problematic in several ways. First, it creates a problem for local grid balancing, especially in regions with a high share of solar PV electricity generation. A balanced grid requires electricity supply and demand to be equal at all times. Additionally, it poses a challenge to grid reliability, as well as to capital recovery. Thus, it is operationally important to be able to predict the power output from PV cells in order to prepare for supply intermittency.

In this project, we explore the Nowcast model, which predicts the output power (in kW) from a solar PV cell given an image of the sky. We utilize a novel approach to PV power output prediction and forecasting, applying a transformer architecture with frozen layers and a subset of the available three-year historical data collected to the visual recognition regression problem. We find that this approach works and performs almost as well as the full baseline model which is a CNN architecture.

## 2. Related work

The application of convolutional neural networks (CNNs) to the problem of solar forecasting is a relatively recent phenomenon [4]. We build on research performed in the Environmental Assessment and Optimization (EAO) Lab group at Stanford [5], [6], using their baseline solar output forecasting models (known as the SUNSET models) for the Nowcast [7] and Forecast [8], [9]. The use of transformers for computer vision tasks is also quite recent [10]. Dosovitskiy et. al. (2020) [11] apply a pure transformer architecture to a sequence of image patches and showed improved performance on small-sized classification tasks when pre-trained on large datasets. In this project, we will explore the performance of transformers on the Nowcast and Forecast problems, adapting the details of the Vision Transformer [11][12] to our use case.

## 3. Data

The data set consists of one year (2017) of video recordings of the sky above a PV solar array, pre-processed to snapshot minutely, high resolution images of daytime sky, each of size 2048 × 2048 pixels. Images are then downsampled to a desired size (64 × 64, 128 × 128, 256 × 256, etc.). The sky images are collected in-house via a fisheye camera positioned on the rooftop of Stanford's Green Earth Sciences Building while the 30.1 kilowatt (kW) polycrystalline PV array is positioned on the roof of the Jen-Hsun Huang Engineering building. In conjunction with the sky images, we have the corresponding minutely PV power output data. The minutely snapshot images are then paired with the corresponding PV output data, resulting in a data set consisting of 103,209 image-PV output pairs. The desired prediction from the model is the (non-negative) power output generated by the PV array (in kW).

We use images of resolution 256 × 256 with a 95% − 5% split for training and validation. The data was shuffled before performing the split. This was done to prevent the model from learning temporal patterns in the data, i.e. we did not want any one prediction to depend heavily on the prediction for the previous image. This was important since images may not be in any particular order at test time and we wanted the model to perform well in these cases.

The test data set consists of 20 days of image-PV output pairs (13,689 such pairs). These images have a lower resolution of 64 × 64. 10 of these days are classified as sunny and the other 10 as cloudy by visual inspection. This is the same test dataset used by the baseline SUNSET model.

We used the pre-trained HuggingFace ViT model for this project. For feature extraction, we utilise the pre-trained feature extractor from the same model. This feature extractor has been pre-trained on the ImageNet-21k dataset at a resolution of 224 × 224. Our dataset is rescaled from 256 × 256 during the feature extraction. Sample visualizations of the extracted features are presented in figure 5.1.

## 4. Methods

Within the past year, transformer architectures have generalized from applications in the natural language processing field to computer vision. This motivated us to assess the performance of the vision transformer to our solar power forecasting problem. Initially, we considered a hybrid CNN-transformer approach. However, Vaswani et al (2017) [13] show that the transformer architecture is able to utilise the attention mechanism as an alternative to convolutions. We therefore applied a pure transformer architecture to the solar forecasting problem. The existing model from the EAO lab group (SUNSET) relies purely on CNN architectures to predict PV power output. We utilize their data and setup implementation, coupling with the transformer architecture of Dosovitskiy et al. (2020) [11]. We modify this for our regression problem by using a mean-squared error loss on a one-class classification. A schematic of the ViT architecture we use is shown in Figure 1.

### 4.1. Baseline model

The architecture of the baseline SUNSET Nowcast model to which we compare our results consists of two convolution layers (with batch normalization and max pooling following each convolution layer), followed by two fully-connected layers. Figure 2 depicts the SUNSET model architecture and workflow.

| SUNSET (64x64) RMSE |
|---|
| Sunny: 0.52 |
| Cloudy: 3.38 |
| Overall: 2.36 |

Table 1. *SUNSET Nowcast baseline root mean squared error (RMSE) values.*

Adam is used as the optimizer, and MSE (mean squared error) is used to force the output to match the PV power labels. With the architecture described above applied to sky-images of 64x64 pixels, we observe baseline performance results as shown in Table 1. The best RMSE achieved on our test set of ten days (five cloudy and five sunny) is 2.36.

### 4.2. Transformer model

As mentioned before, we utilize HuggingFace Vision Transformer (ViT) model with pre-trained weights [14] for our implementation. This model has been pre-trained on the ImageNet-21k dataset ($\approx$ 14 million images) at a resolution of 224 × 224 and finetuned on the ImageNet 2012 dataset ($\approx$ 1 million images) at the same resolution. The model we used took in patches of resolution 16 × 16 embedded linearly. Once again, we used Adam as the optimizer and MSE for the loss. The performance from our best model is presented in Table 2.

| ViT (64x64) RMSE |
|---|
| Sunny: 2.43 |
| Cloudy: 4.19 |
| Overall: 3.38 |

Table 2. *Best ViT root mean squared error (RMSE) values.*

## 5. Experiments

A pilot experiment using a single day of image-output PV data (654 image-output pairs) resulted in a root mean-squared error (RMSE) of 2.61 (see Table 3) on the same test set utilized for the baseline SUNSET Nowcast model. For this experiment we used the aforementioned pre-trained ViT model with no hyperparameter tuning. This indicated that transformers are well-suited to our PV power output regression task. To further corroborate this observation, we performed several experiments on the entire data set.

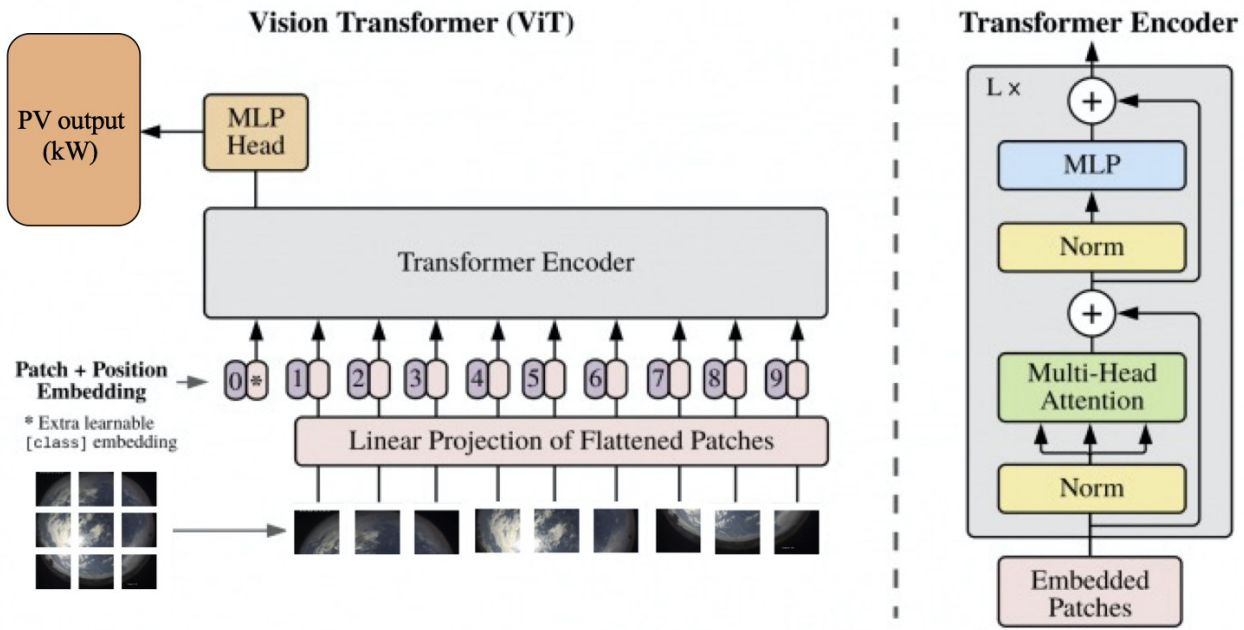

Figure 1. *Vision transformer architecture from [11] adapted to the solar forecasting problem*

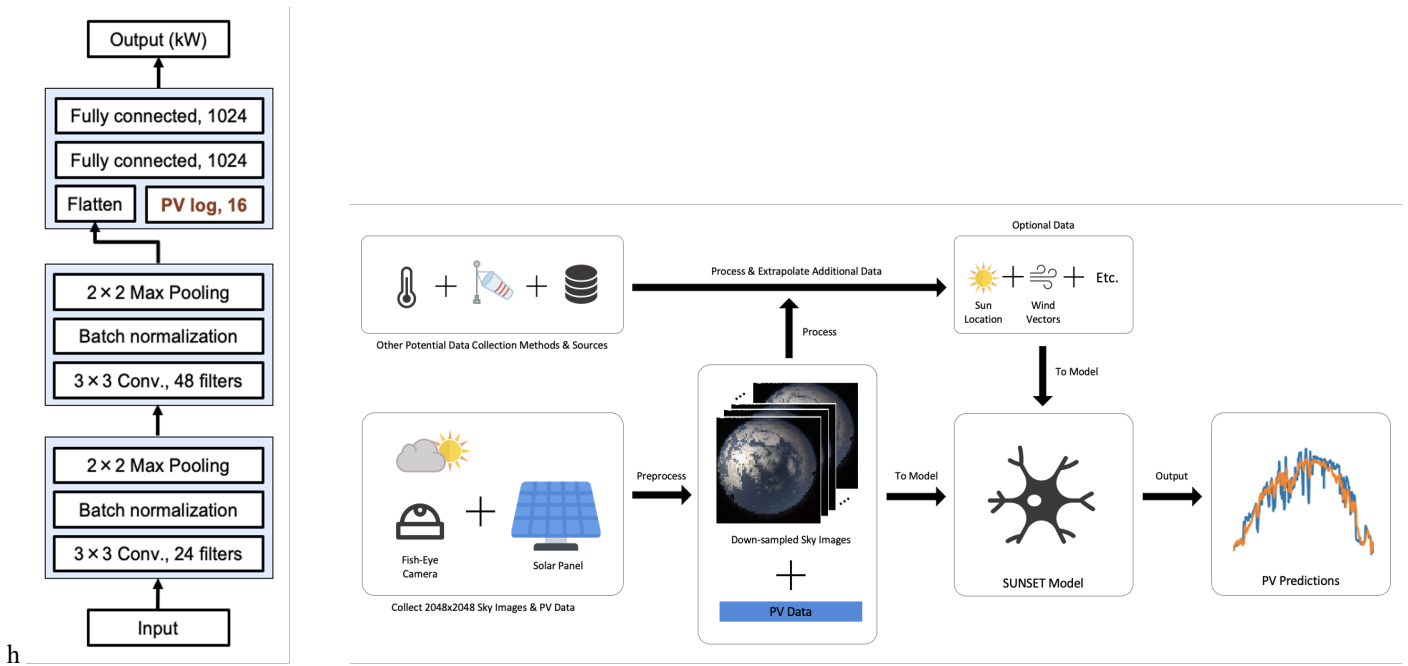

Figure 2. SUNSET baseline model architecture *(left)* and workflow *(right)*

| **ViT (224x224) RMSE** |
| --- |
| Overall: 2.61 |

Table 3. *ViT Nowcast baseline root mean squared error (RMSE) values.*

In the first experiment, we trained the same model used in the pilot study on our entire dataset to verify whether ad-ditional data improved the performance of the model. To reduce the training time, we performed all subsequent experiments by freezing the weights on all the layers of the model except for the final layer normalization, pooler layer and fully connected layer. This allowed us to perform multiple experiments in which we tuned the learning rate and batch size hyperparameters. The results from these experiments are presented in the following sub-sections.

### 5.1. Feature extraction and maps

Using 654 samples of images sized at 64 × 64 from the data set and split 90% − 5% into training and validation sets, we extract features from the image data with the ViT pre-trained model.

Figure 5.1 shows sample feature maps of extracted image features from the validation set.

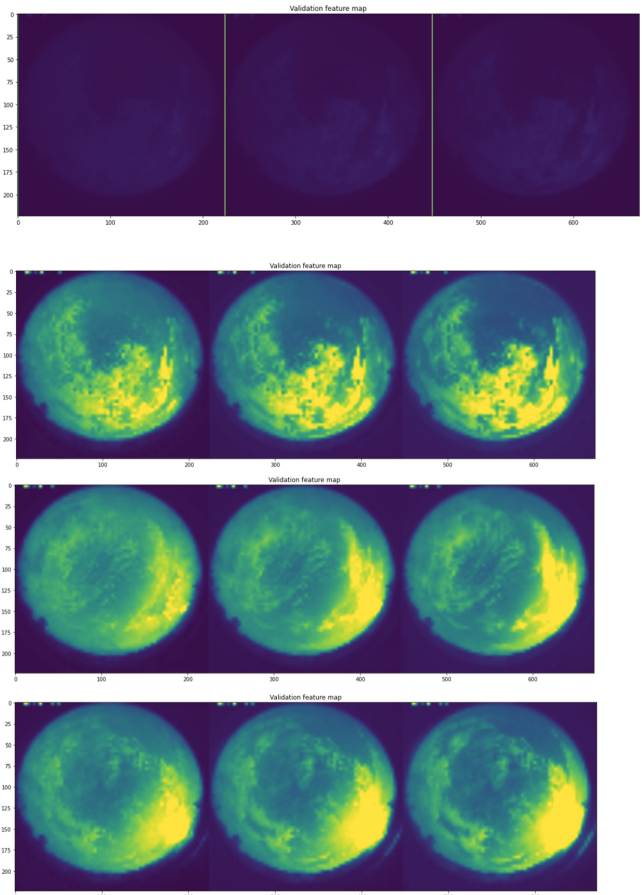

Figure 3. Image feature maps

### 5.2. Training with Full Dataset: Warm Start

In the first experiment, we trained the pre-trained ViT model on the full dataset using the pretrained weights, a learning rate of $1e-5$, and batch size of 64. The experiment was testing a warm-start where all of the pretrained weights were loaded with no weights being frozen. Thus, all weights were further trained on the dataset. Figure 5.2 depicts the training and validation losses, showing great improves and starts to attain a minimum around epoch 15.

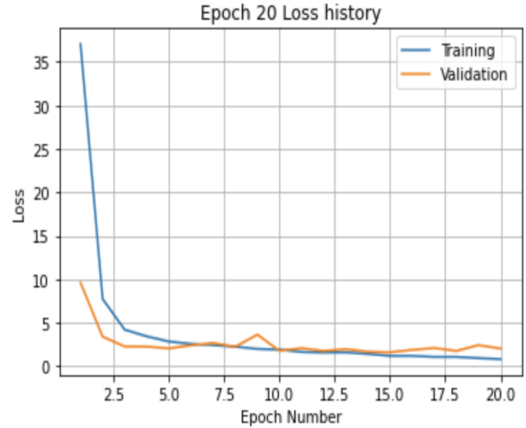

Figure 4. Training loss history.

The model performance is presented in 4.

| ViT (no frozen layers) test RMSE |
| --- |
| Sunny:   XX |
| Cloudy:   XX |
| Overall: 2.61 |

Table 4. *ViT model RMSE values when all layers are trained on full dataset.*

### 5.3. Hyperparameter Tuning

The next experiment we performed was hyperparameter tuning. Due to the time intensive nature of training all the layers of the model, we froze the weights on all the layers of the model except for the final layer normalization, pooler layer and fully connected layer for these experiments. We then performed hyperparameter tuning on the learning rate and the batch size. We trained models over a combination of learning rates of $[1e-6, 5e-6, 1e-5, 5e-5, 1e-4]$ and batch sizes of $[10, 64]$.

Figure 5.3 depicts the training and validation losses for these experiments. Table

The model performance for the best model is presented in figure 5. The performance of this model on the test set is visualized in 5.3. The blue curve represents the ground truth and the orange curve represents the model prediction.

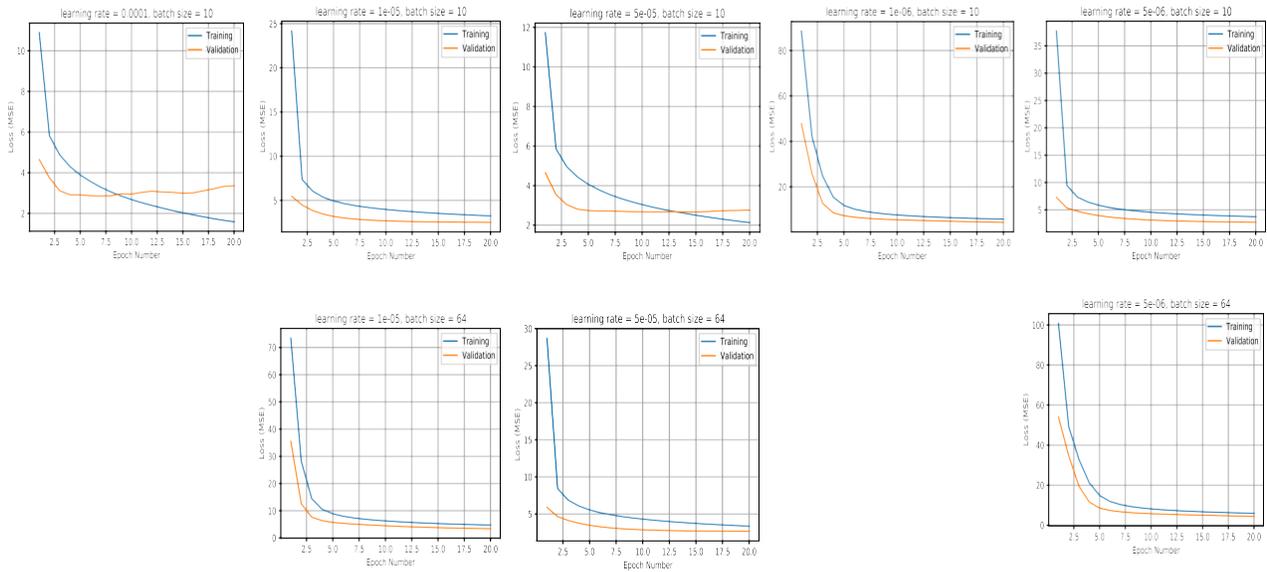

Figure 5. *Training loss on experiments varying learning rates (*$1e^{-6} : 1e^{-4}$*) and batch size (*$[10, 64]$*)*

| Learning rate | Batch size | RMSE: sunny | RMSE: cloudy | RMSE_overall | Final loss |
|---|---|---|---|---|---|
| 1E-06 | 10 | 2.75 | 4.21 | 3.52 | 12.37 |
| 5E-06 | 10 | 2.55 | 4.19 | 3.43 | 11.75 |
| 5E-05 | 10 | 2.62 | 4.01 | 3.35 | 11.21 |
| 5E-05 | 10 | 2.54 | 4.25 | 3.45 | 11.92 |
| 1E-04 | 10 | 2.43 | 4.19 | 3.38 | 11.43 |
| 5E-06 | 64 | 2.78 | 4.26 | 3.56 | 12.67 |
| 5E-05 | 64 | 2.52 | 4.21 | 3.42 | 11.72 |
| 5E-05 | 64 | 3.03 | 4.45 | 3.77 | 14.19 |

Figure 6. Hyperparameter tuning results

| ViT (frozen layers) test RMSE |
|---|
| Sunny: 2.62 |
| Cloudy: 4.01 |
| Overall: 3.35 |

Table 5. *ViT model test RMSE values trained on full dataset when most layers are frozen. Learning rate = $1e-5$ and batch size = 10.*

## 6. Conclusion

Additionally, we may incorporate a CNN component to see if this will improve the RMSE. Finally, we hope to expand this model to our SUNSET Forecast problem.

## 7. Acknowledgments

We would like to acknowledge Adam Brandt, Yuhao Nie, and Alexandre Matton for their support.

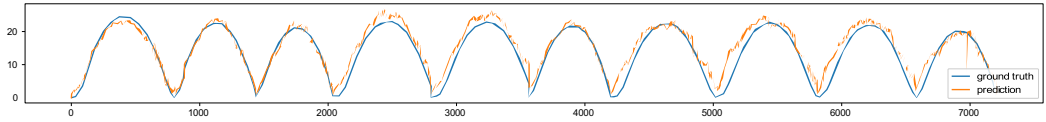
Figure 7. Cloudy

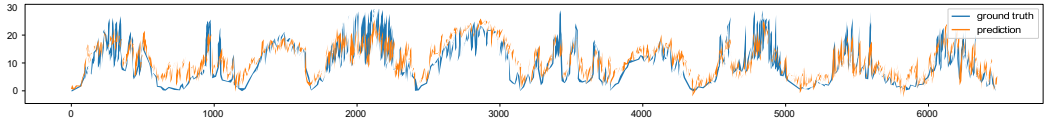
Figure 8. Sunny

Figure 9. Model prediction and ground truth solar PV power output